\documentclass[twocolumn,showpacs,amsmath,amssymb,prl]{revtex4}

\usepackage{graphicx}
\usepackage{dcolumn}
\usepackage{bm}
\usepackage{psfrag}

\usepackage{color}
\definecolor{red}{rgb}{0.8,0.0,0.0}

\definecolor{blue}{rgb}{0.0,0.0,0.8}

\def\up{\uparrow}
\def\dn{\downarrow}

\begin{document}

\title{Non-adiabatic two-parameter charge and spin pumping in a quantum dot}

\author{Matthias Braun}
\affiliation{Institute of Theoretical Physics C, RWTH Aachen University, D-52056 Aachen, Germany}
\author{Guido Burkard}
\affiliation{Institute of Theoretical Physics C, RWTH Aachen University, D-52056 Aachen, Germany}

\date{\today}
\begin{abstract}
We study DC charge and spin transport through a weakly coupled quantum dot, driven by a non-adiabatic periodic change of system parameters.  We generalize the model of Tien and Gordon to simultaneously oscillating voltages and tunnel couplings. When applying our general result to the two-parameter charge pumping in quantum dots, we find interference effects between the oscillations of the voltage and tunnel couplings.  Furthermore, we discuss the possibility to electrically pump a spin current in presence of a static magnetic field.
\end{abstract}

\maketitle

A periodic perturbation of the parameters that determine a quantum dot and 
its coupling to external leads can lead to an electric (DC) current from one
of the attached leads to the other.  This phenomenon is known as charge pumping.
For perturbations that are slower than the characteristic charge dynamics of the quantum dot,
the pumping process is adiabatic  \cite{Thouless1983,Brouwer1998,Splettstoesser2005,Pothier1992,Switkes1999,Dicarlo2003,Watson2003,Altshuler1999}. 
The relevant time scale of the charge dynamics is typically given by the tunnel rates
betweed dot and leads. A remarkable property of adiabatic pumping is that the pumped charge
is independent of any details of the pumping cycle, making it possible to realize 
a current standard for metrology.

Fast perturbations, with frequencies exceeding the tunnel rates, 
can still lead to pumping effects which in this case are non-adiabatic. 
Non-adiabatic pumping in quantum dots has broad applications, reaching from photovoltaic power generation \cite{Nozik2002} to fundamental studies of fast manipulations of quantum systems, as required for example in quantum information processing. The driving force behind non-adiabatic 
pumping is the absorption of quantized photon energy. Therefore, non-adiabatic pumping \cite{vanderWiel2001, Dovinos2005, Kouwenhoven1994,Kouwenhoven1994-2, Qin2001,Shin2006,Stoof1996, Hazelzet2001} is often studied as a side-effect of boson-assisted tunneling \cite{Oosterkamp1997, Blick1995}.

\begin{figure}[t]
\includegraphics[width=.8\columnwidth]{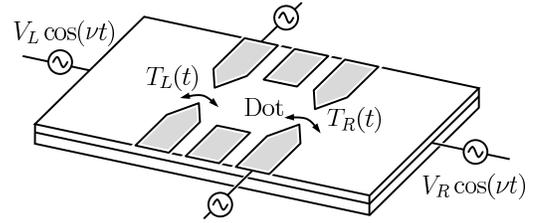}
\caption{\label{fig:quantumdot}
Typical realization of a quantum dot pump in a 2-dimensional electron gas patterend with gates. 
By a sufficiently fast periodic perturbation of voltages $V_r \cos (\nu t)$ or tunnel couplings $T_{rk}(t) =\bar{T}_{rk}[1+\alpha_r \cos(\nu t+\eta_r)]$ at the left ($r=L$) or right ($r=R$) contact, non-adiabatic processes can pump charge from one to the other lead. 
}
\end{figure}

Although non-adiabatic pumping is observed in different quantum dot realizations, such as carbon nanotubes \cite{Shin2006}, or self-assembled dots, the most common realization of a quantum dot is in a two-dimensional electron gas. By charging gates, the electron gas beneath can be repelled, and a quantum dot and tunnel barriers can be formed, see Fig.~\ref{fig:quantumdot}. This structure has the advantage that many system parameters are controllable. Operated at frequencies of $1-100$ GHz, currents of the order pA to nA can be generated.

%
%
%
%
%
%

In non-adiabatic pumping experiments, it is usually assumed that pumping originates 
from an oscillating voltage of the leads or the energy of the electronic levels in the quantum dot 
\cite{vanderWiel2001, Dovinos2005, Kouwenhoven1994,Kouwenhoven1994-2, Qin2001,Tien1963,Stoof1996, Hazelzet2001}, while a variation of the tunnel barrier height is less discussed (see, however, Refs.~\onlinecite{Mahmoodian2006, Moldoveanu2007}). At this point it should be noted that in particular in 2DEG quantum dots, tunnel couplings are exponentially sensitive to voltage changes and therefore 
an oscillating tunnel-barrier may well be the dominant source of photon energy.

In the following, we present an extension of the well-known model by Tien and Gordon \cite{Tien1963} that also includes tunnel barrier oscillations.  We find that barrier-induced pumping may explain several experiments  \cite{vanderWiel2001, Dovinos2005, Kouwenhoven1994,Kouwenhoven1994-2, Qin2001}.  Furthermore, in the case of two oscillating parameters, one can observe interference between the different sources of pumping, which possibly explains the observed asymmetry of forward and backward currents in Ref.~\onlinecite{Dovinos2005}.

As a model system, we consider a single-level quantum dot contacted to two leads, described by the Hamiltonian
\begin{eqnarray}\label{dots}
H(t)&=&\sum_{\sigma} \varepsilon\, a^\dag_{\sigma}a^{}_{\sigma}
+ U a^\dag_{\up}a^{}_{\up}a^\dag_{\dn}a^{}_{\dn}+\sum_{r,k,\sigma} \varepsilon^{}_{rk}(t) c^\dag_{rk\sigma} c^{}_{rk\sigma}\nonumber\\
&&\!\!\!\!\!\!\!\!+\sum_{r,k,\sigma} T_{rk}(t) c^\dag_{rk\sigma} a^{}_{\sigma}+ T_{rk}^\star(t) a^{^\dag}_{\sigma}c^{}_{rk\sigma}\,. 
\end{eqnarray}
The fermionic operators $a^\dag_{\sigma}/a^{}_{\sigma}$ create/annihilate electrons with spin $\sigma$ on the dot, while the operators $c^\dag_{r k\sigma}/c^{}_{r k\sigma}$ act on electrons with orbital quantum number $k$ in the left $(r=L)$ and right $(r=R)$ lead. 

Due to its low electrostatic capacity, we assume that double occupation of the dot is forbidden as the associated charging energy $U$ exceed all other energy scales. The third term of the Hamiltonian models the contacting leads on the left and right side $(r=L/R)$. The two lead reservoirs are assumed to be characterized by the same Fermi distribution $f(\omega)$ as long as no static bias voltage or temerature gradiend is applied. The last part of the Hamiltonian  describes spin-conserving tunneling.

We consider harmonically oscillating voltages at the left and right electrode, leading to a time-dependence of the electron energy $\varepsilon_{rk}(t)=\bar{\varepsilon}_{rk}+e V_r \cos(\nu t)$ in the respective lead. Furthermore, we allow for a time-dependent tunneling amplitude $T_{rk}(t) =\bar{T}_{rk}[1+\alpha_r \cos(\nu t+\eta_r)]$ \cite{note1,note2}.
The relative phase of the tunneling-amplitude and voltage oscillations on the same side is given by $\eta_r$. The tunnel coupling leads to an (time-averaged) intrinsic linewidth $\gamma_{r}(\omega)=2\pi\sum_k |\bar{T}_{rk}|^2 \delta(\varepsilon_{rk}-\omega)$ of the quantum dot levels. For simplicity, we will neglect the energy dependence of the linewidth in the following.

After tracing out the degrees of freedom of the non-interacting leads, the quantum dot is described by a reduced density matrix.
As in this simple model the reduced density matrix is always diagonal, ensured by particle and spin conservation, it is convenient to express it as a vector
${\bf p}=  (p_0,p_\uparrow,p_\downarrow)^{\text{T}}$. 
The elements $p_\chi$ describe he probability to find the dot in the state empty $(\chi=0)$, or occupied with one electron with spin $\sigma$ $(\chi=\sigma)$.

The time evolution of the density matrix is given by the Master equation
$\frac{d}{dt}{\bf p}\left(t\right)=\int_{-\infty}^{t}dt'\bm{\Gamma} \left(t,t';\varepsilon\right){\bf p}\left(t'\right)$ where the elements $\Gamma_{\chi,\chi^\prime}\left(t,t';\varepsilon\right)$ of the kernel describe the transition rates from the state $\chi'$ at time $t'$ to a state $\chi$ at time $t$.
If the system parameters change faster than the typical time scale of the system evolution, one can assume that the quantum dot density matrix adapts a steady state $\bar{\bf p}={\bf p}(t^\prime)={\bf p}(t)$, which satisfies 
$0= \bar{\bm{\Gamma}} \cdot \bar{\bf p}$,
with the end-time averaged kernel $\bar{\bm{\Gamma}}(\varepsilon)=\int_{0}^{2\pi} d(\nu t)/2\pi \int_{-\infty}^{t}dt'\,\bm{\Gamma} \left(t,t';\varepsilon\right)$
In contrast to more sophisticated approaches like Floquet theory \cite{Stafford1996, Platero2003, Kohler2004}, the approximation of separating time scales \cite{Stoof1996,Hazelzet2001} covers only the highly non-adiabatic regime. However, as in the weak-coupling regime the charge dynamics of the quantum dot  slows down exponentially in $\gamma_r$, this approximation is reliable even if the oscillation frequency $\nu$ exceeds $\gamma_r/\hbar$  by less than an order of magnitude.

In the following we will only discuss a weakly-coupled quantum dot, where an expansion of the kernel in lowest, i.e., first order $\gamma_r$ is a reasonable approximation. In this regime, the kernel can be decomposed into a left and right part $\bm{\Gamma} \left(t,t';\varepsilon\right)=\bm{\Gamma}^L \left(t,t';\varepsilon\right)+\bm{\Gamma}^R \left(t,t';\varepsilon\right)$, which contain only tunneling processes from and to the respective lead.

A typical expression for a rate is
\begin{eqnarray}
\label{rate}
\Gamma^r_{0,\sigma}\left(t,t^\prime;\varepsilon\right)&=&
\sum_k T_{rk}(t)T^\star_{rk}(t^\prime) f(\bar{\varepsilon}_k) \times \nonumber \\
&&e^{-i\int_{t^\prime}^t \!d\tau [\varepsilon-\varepsilon_k(\tau)+i0^+]}+c.c.\,.
\end{eqnarray}
As the phase term depends only on the voltage difference between dot and lead, a time dependence of the quantum dot level $\varepsilon\rightarrow \varepsilon(t)$ is not qualitatively different from a time dependent voltage applied to the lead(s). Note that since both the chemical potential and the lead particle energy $\varepsilon_k$ are shifted by an applied voltage, the argument of the Fermi function in Eq.~(\ref{rate}) is $\bar{\varepsilon}_k$ not influenced by voltage.

After the phase averageing, the first-order rates transform into
\begin{eqnarray}
\label{eq:mainrates}
\bar{\Gamma}^r_{\!\chi,\chi^\prime}\left(\varepsilon\right)=
\!\!\!\!\!\sum_{n=-\infty}^\infty\!
\left| J_n+\frac{\alpha_r e^{i\eta_r}}{2} J_{n+1}+\frac{\alpha_r e^{-i\eta_r}}{2} J_{n-1}\right|^2\times\nonumber \\
\Gamma^{0,r}_{\!\chi,\chi^\prime}\left(\varepsilon+n\,\hbar\nu\right)\,,\quad
\end{eqnarray}
where $J_n\equiv J_n(eV_r/\hbar\nu)$ is the Bessel function of the first kind, and $\Gamma^{0,r}_{\chi,\chi^\prime}(\varepsilon)$ is the well-known Golden Rule rate for the time-independent problem. With these modified rates the static occupation probabilities of the quantum dot as well as the current can be calculated \cite{Thielmann2003}. Equation (\ref{eq:mainrates}) generalizes the result of Tien and Gordon \cite{Tien1963} for an oscillating bias voltage, to take into account an oscillating barrier strength. It is worth mentioning that Eq.~(\ref{eq:mainrates}), with modified Golden Rule rates, also holds in the case of a metallic island instead of a quantum dot. 

In the following, we use this general result to discuss charge pumping and spin pumping in quantum dots.
Let us first focus on the situation where an electrical DC-current is generated by the oscillation of one or more system parameters, in the absence of a bias voltage \cite{note3}. For functional clarity we consider only small system parameter changes  $\alpha_r \ll 1$ and $e V_r/\hbar\nu\ll 1$, i.e., we calculate only the quadratic response to the system parameter change. The total current through the quantum dot can be written as 
\begin{eqnarray}
\label{eq:totalcurrent}
I=I_{V_L^2}+I_{\alpha_L^2}+I_{V_L\cdot\alpha_L}-(L\rightarrow R)\,.
\end{eqnarray}

\begin{figure}
\includegraphics[width=.8\columnwidth]{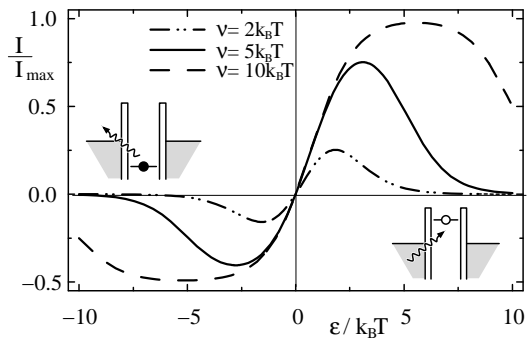}
\caption{\label{fig:leftbarrier}
Pumped particle current by one weakly oscillating system parameter with frequency $\nu$ on the left side, either voltage $V_L$ or tunnel amplitude $\alpha_L$. The current is driven by the absorption of a photon energy  $\hbar \nu$ when tunneling from or to the left lead.  The maximal reachable pumped current is given by $I_{\rm max}=(e V_L/\hbar\nu)^2 I_0$ and  $I_{\rm max}= \alpha_L^2 I_0$ respectively.}
\end{figure}

If the tunnel amplitudes are constant in time $(\alpha_r=0)$, an applied AC-bias voltage at the quantum dot structure generates the pumped current
\begin{eqnarray}
\label{eq:leftvoltage}
I&=& I_{V_L^2}-I_{V_R^2}\\
&=&I_0 
\frac{(eV_L)^2-(eV_R)^2}{1+f(\varepsilon)}
\frac{f(\varepsilon+\nu)-2 f(\varepsilon)+f(\varepsilon-\nu)}{\nu^2}\,,\nonumber 
\end{eqnarray}
with $ I_0=e/\hbar\cdot\gamma_L \gamma_R/(\gamma_L+\gamma_R) $. As in real experiments 
the capacitance of the left and right tunnel barrier always differ, the voltage drop over the the two tunnel barriers will be asymmetric, $V_L\neq V_R$, and a net current is pumped. 
For oscillation frequencies $\hbar \nu \gg k_{\rm B}T$ larger than temperature, the maximally pumped current scales as $(e V_r/\hbar \nu)^2<1$. For $\hbar \nu \ll k_{\rm B}T$, the second fraction in Eq.~(\ref{eq:leftvoltage}) can be approximated by the second derivative of the Fermi function, and the pumped current scales as $(e V_r/k_BT)^2$.  In Fig.~\ref{fig:leftbarrier} the current, pumped by an oscillation of the left voltage only, is plotted in units of $I_{\rm max}=I_0 (e V_L/\hbar \nu)^2$ as function of the dot level position $\varepsilon$. 
%
%
%
%
%
%
%
%

The mechanism behind the pumping lies in the possibility of the electrons to absorb or emit a photon energy $\hbar \nu$ when tunneling from or to the left lead as discussed in Ref. \onlinecite{Kouwenhoven1994}. If the quantum dot level lies above the Fermi energy $(\varepsilon>0)$, a left-lead electron can absorb a photon, and tunnel onto the dot.
For $(\varepsilon<0)$, the absorption of a photon enables the dot electron to tunnel to the left lead.  A successive tunnel event to (from) the right lead creates a positive (negative) particle current. The strength of these currents differ by a factor up to 2, originating from the spin degeneracy of the initial and final state of the tunneling event. This relative factor, however, appears only in experiments with quantum dots \cite{Shin2006}, i.e., in the case of a discrete energy spectrum rather than a continuous spectrum as in the case of metallic islands \cite{vanderWiel2001, Dovinos2005, Kouwenhoven1994, Kouwenhoven1994-2, Qin2001}.

In absence of an AC-voltage applied at the leads $(V_r=0)$, a current $I=I_{\alpha_L^2}-I_{\alpha_R^2}$ can also be driven by an AC-signal on one of the gates leading to an oscillation of the left and/or right tunneling amplitudes. The functional form of the current is also given by Eq.~(\ref{eq:leftvoltage}) where $(eV_r)^2$ is replaced by $(\alpha_r \nu)^2$. Therefore Fig.~\ref{fig:leftbarrier} also describes the pumped current pumped by a tunnel-barrier oscillation, with the current scale $I_{\rm max}= \alpha_L^2 I_0$. 
This similarity of the two results is a direct consequence of the closely related physical origin of the current. This similarity also raises the question, if the observed pumped current in the experiments  \cite{vanderWiel2001, Dovinos2005, Kouwenhoven1994, Kouwenhoven1994-2, Qin2001} is driven by an oscillating voltage or by an oscillation of the tunnel barrier.

To discriminate if an oscillating barrier or the oscillating voltage drop causes the current, one needs to look for multi-photon absorption processes. If the voltage amplitude $eV_r$ exceeds the oscillation frequency $\hbar\nu$, multi-photon absorption processes become possible, as observed for example in Ref.~\onlinecite{vanderWiel2001} in the photon-assisted tunneling measurements. 
In contrast to an oscillating voltage, the oscillation of a weak tunnel barrier can lead only to the absorption of a single photon, and not to multi-photon absorption. This is a direct consequence of the lowest-order expansion in the tunneling strength. Two-photon absorption would first be possible in cotunneling events.

\begin{figure}
\includegraphics[width=.8\columnwidth]{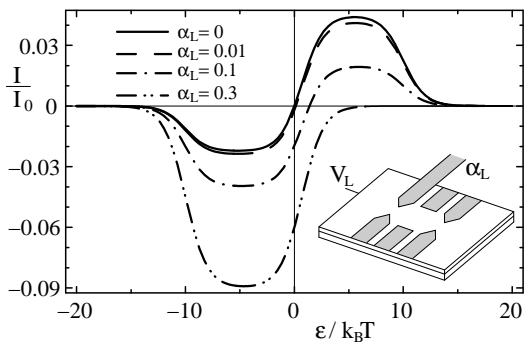}
\caption{\label{fig:twoparameters}
Current pumped by an AC-voltage $eV_L=0.3 \hbar \nu $, and an in-phase $(\eta_L=0)$ tunneling amplitude oscillating $\alpha_L=0,0.01,0.1$, and $0.3$. While the boson-assisted tunneling on the dot gets suppressed, the tunneling out of the quantum dot gets enhanced.}
\end{figure}

Due to  mutual cross capacitances, it is rather unlikely that an experimentally applied voltage to either the gates or at the electrodes will change only one system parameter.  It is more likely that the voltage at the lead as well as the tunnel coupling strength will start to oscillate. As can be  seen in Eq.~(\ref{eq:totalcurrent}), the two pumping processes do not only coexist, but give rise to a mixed two-parameter pumping term
\begin{eqnarray}
\label{eq:twoparameter}
I_{V_L\cdot\alpha_L}=4 I_0 \cos(\eta_L)
\frac{e V_L \alpha_L}{1+f(\varepsilon)}
\frac{f(\varepsilon+\nu)-f(\varepsilon-\nu)}{2 \nu}\,.
\end{eqnarray}
Several differences to the one-parameter pumping currents appear. First, the last term in 
Eq.~(\ref{eq:twoparameter}) resembles the first derivative of the Fermi function. Therefore, especially in the limit of high temperatures $k_{\rm B}T>\hbar\nu$ the maximally possible pumped current $4 I_0 eV_L \alpha_L/{\rm max}[\hbar \nu, k_{\rm B}T]$ can significantly exceed the one-parameter pumping currents. 
Second, the mixed term depends on the relative phase $\eta_L$ of the voltage and tunnel amplitude oscillation on one side. This is an indication of quantum mechanical interference of the two sources of photon energy $\hbar \nu$. In Fig.~\ref{fig:twoparameters} the total current $I$ is plotted for an in-phase oscillating source lead voltage with amplitude $e V_L = 0.3 \hbar \nu$ for increasing tunneling barrier amplitudes $0\leq\alpha_L\leq e V_L$ for $\hbar\nu=10k_{\rm B}T$.
Due to the different numerical prefactors even a tunnel amplitude change of few percent already leads to a noticeable change of the pumped current. Interestingly,  for $\eta_L=0$, which one would expect for an unintentional parameter change, the interference of the two pumping possibilities is constructive if an electron tunnels out of the dot onto the lead and destructive if tunneling from the lead to the dot. Therefore the negative current is enhanced, while the positive one is suppressed.
In the pumping experiment by Dovinos and Williams \cite{Dovinos2005}, an asymmetry in the forward and backward pumping direction was observed, which can be an indication of such a two-parameter pumping situation.

%
%
%
%
%
%
%
%

\begin{figure}
\includegraphics[width=.8\columnwidth]{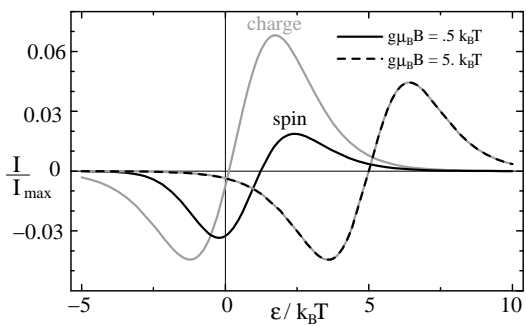}
\caption{\label{fig:spin}
Particle current (gray) and spin current $I_s$ (black) pumped by oscillating tunnel barrier at the left contact for $\hbar \nu = k_B T$. For lower magnetic fields ($g\mu_{\rm B}B=0.5 k_{\rm B}T$) a net spin current can be generated even in absence of charge current. For high magnetic field $g\mu_{\rm B}B=5 k_{\rm B}T$, when only one spin state in the quantum dot participates to transport, the charge current is fully polarized. }
\end{figure}

By applying an additional static magnetic field, it is also possible to non-adiabatically pump a spin current. Let us consider the case where current is driven only by an oscillation of the left tunnel barrier.  The particle current carried by electrons with spin $\sigma$ only is for $\hbar\nu\ll k_{\rm B}T$ given by
\begin{eqnarray}
\label{eq:spin}
I_{\alpha_L^2}^{\sigma}=(\nu \alpha_L^2)\frac{I_0}{2} \frac{1-f(\varepsilon_{\bar{\sigma}})}{1+f(\varepsilon_{\bar{\sigma}})f(\varepsilon_{\sigma})}
\left.\frac{d^2f(\varepsilon)}{d\,\varepsilon^2}\right|_{\varepsilon=\varepsilon_\sigma}\,.
\end{eqnarray}
For currents of order pA to nA the dwell time of the electrons on the dot is of order ns to ps. As the spin relaxation time significantly exceed this dwell time, spin-flip processes can be neglected.
In Fig.~\ref{fig:spin}, the particle current (gray) and the spin current  $I_s=I_{\alpha_L^2}^{\sigma}-I_{\alpha_L^2}^{\bar{\sigma}}$ (black) is plotted for different magnetic fields in units of $I_{\rm max}= \alpha_L^2 I_0$. For the magnetic field $g\mu_{\rm B}B=0.5k_{\rm B}T$ one can observe that around the Fermi energy the charge/particle current shows a node while the spin current bears a maximum. Therefore, a non-adiabatic one-parameter pump can drive a pure spin current without any charge current. For a larger magnetic field, $g\mu_{\rm B}B=5 k_{\rm B}T$ for example, only one spin component still participates to transport, therefore the charge and spin currents become equal.
In contrast to spin pumping schemes relying on electron spin resonance in quantum dots \cite{EngelLoss2001,Koppens2006}, this proposal does not require a strong and fast oscillating magnetic field. Instead, the current is purely driven by an oscillating electric field, and only a static magnetic field is needed for breaking the spin symmetry, in close analogy to the adiabatic pumping case \cite{Watson2003}.

%
%
%
%
%
%

In conclusion, we have analyzed an extension of the Tien and Gordon model \cite{Tien1963}, taking into account simultaneously oscillating tunneling barriers and voltages. We have used this general result to discuss two-parameter charge pumping in quantum dots. Thereby we observed a quantum interference of the tunneling transitions driven by the different pumping parameters. Furthermore, we have discussed the possibility to electrically drive a spin current in absence of a charge current.

This work was supported by DFG SPP Spintronics.


\begin{thebibliography}{24}
\bibitem{note2}
The model implies, that leads and gates individually follow adiabatically the applied voltages, i.e. $\nu$ must be slow compred to the plasma frequency/RC-response time.
\bibitem{note3}
With an applied bias voltage, Eq.~(\ref{eq:mainrates}) could also be used to discuss boson-assisted tunneling in the presence of oscillating barriers.
\end{thebibliography}
\end{document}